\documentstyle[aps,tighten,preprint,epsfig]{revtex}
\begin{document}
\title{Study of Nucleon Resonances with Double Polarization
Observables of Pion Photoproduction}

\author{D. Dutta$^a$, H. Gao$^a$, T.-S. H. Lee$^b$ }

\address{$^a$ Laboratory for Nuclear Science and
Department of Physics, Massachusetts Institute of Technology,
, Cambridge, Massachusetts 02139}
\address{$^b$ Physics Division, Argonne National Laboratory, Argonne, Illinois 60439.}
\maketitle

\begin{abstract}
The role of the nucleon resonances in the double polarization 
observables of pion photoproduction is investigated by using the 
resonance parameters predicted by Capstick and Roberts.
As an example, we show that the not-well-determined two-star
resonance N$^{-}_{3/2}$(1960) can be examined  by
performing experiments on beam-recoil polarization at large angles.
\end{abstract}  

\section{INTRODUCTION}
Constituent quark models have been widely used to 
predict the properties of baryon resonances. However,
a large number of resonance states which appear in the  
constituent quark models have not been observed 
in pion-nucleon scattering. 
It has been argued~\cite{koniuk80} that such
``missing resonances'' either decouple from the pion-nucleon 
channel or are masked by 
the neighboring resonances which have large $\pi N$ decay widths.
This has led to suggestions that other reaction channels 
such as vector meson productions\cite{ZLB98,ZDGS99,oh2001}
 might be more sensitive to these missing states. However, all 
possibilities have not been exhausted in the $\pi$~N channel. One such
 possibility  is to consider  
the polarization observables. While efforts\cite{exp} are being
made to measure the single polarizations, it will be interesting to
explore double polarizations which could be more
effective in revealing  some of the 
missing resonances through the interference between
the resonant and background(non-resonant) amplitudes.
Investigations of these double polarization observables is timely
and topical as they coincide with the availability 
of new opportunities to 
measure them at Jefferson Lab and other laboratories.    

As a step in this direction, we will explore in this paper how the 
double polarization observables of pion 
photoproduction can be used to investigate nucleon
resonances in the higher energy region( 1.6 $ < E < $  2.1 GeV) 
in which a lot of missing resonances have been suggested. 
In this energy region, a complete dynamical
approach to pion photoproduction, 
such as that  developed\cite{SL96} in the investigation
of the $\Delta$ resonance, is still being developed. The main difficulty
is due to the large number of resonance states and the complexities
of the couplings between several meson-baryon channels.
For our present very limited purpose, we will introduce some
rather drastic simplifications
in using a formulation which is a direct extension of the
dynamical formulation of Ref.\cite{SL96}. This will then allow us to
address the question concerning the missing resonances by 
relating the resonant amplitude directly to the predictions from
a constituent quark model. As an example, we will consider the
predictions by Capstick and Roberts\cite{Caps92,CR94}.

In section II, we will give explicitly the formulation used in
our calculations.  Our findings will be discussed in section III.

\section{Formulation}
The differential cross section of $\gamma N \rightarrow \pi N$
in the center of mass (c.m.) frame can be written as,
\begin{eqnarray}
\label{amps}
\frac{d\sigma}{d\Omega}=\frac{1}{4}\sum_{\lambda_\gamma}\sum_{m_f  m_i}
| A_{mf,mi,\lambda_\gamma}({\bf q},{\bf k})|^2,
\end{eqnarray}
where ${\bf k}$ and ${\bf q}$ are respectively the momenta of
the incident photon and outgoing pion, and $m_i$ , $m_f$,
and $\lambda_\gamma$   are the
spin projections of the initial nucleon, final nucleon, and
incoming photon, respectively.
For a total invariant mass $W=\sqrt{s}$, the momenta 
$k = \mid {\bf k}\mid$ and $q=\mid {\bf q}\mid$ are
\begin{eqnarray}
k&=&\frac{1}{2\sqrt{s}}[s-M_N^2], \nonumber \\
q&=&\frac{1}{2\sqrt{s}}[(s-M_N^2-M_\pi^2)^2-4M_N^2M_\pi^2]^{1/2},\nonumber
\end{eqnarray}
where $M_N$ and $M_\pi$ are the masses for the nucleon
and pion respectively. The photon laboratory energy $E_\gamma$ is
\begin{eqnarray}
E_\gamma=\frac{1}{2M_N}[s-M_N^2]
\end{eqnarray}

Following the formulation of Ref.~\cite{oh2001}, we write the
pion photoproduction amplitude as
\begin{eqnarray}
\label{ampeq}
 A_{mf,mi,\lambda_\gamma}({\bf q},{\bf k})
= A^{(b.g.)}_{mf,mi,\lambda_\gamma}({\bf q},{\bf k})
+ \sum_{N^*} [A^{N^*}_{m_f,m_i,\lambda_\gamma}({\bf q},{\bf k})].
\end{eqnarray}
In this work, we assume that 
the total amplitude $A({\bf q},{\bf k})$ of Eq.(3)  can be calculated 
from the multipole amplitude 
generated by the SAID program\cite{said}. These empirical
amplitudes are obtained by the following procedure.
For each partial wave, the multipole  amplitude is written
 as a sum of the conventional Born term, vector-meson-exchange, and 
 a many-parameter phenomenological term. The amplitudes are unitarized
by using the empirical $\pi N$ phase shifts which are also available
in the SAID program. 
The empirical pion photoproduction
 multipole amplitude are then obtained by adjusting
the parameters of the phenomenological term to 
perform $\chi^2$-fits
to all of the existing data of $\gamma p \rightarrow \pi^+ n$,
$\gamma n\rightarrow \pi^- p$ and $\gamma p \rightarrow \pi^0 p$
reactions from threshold up to invariant mass W=2.1 GeV. In this
work we use the SM01 solutions\cite{said} 
from their  energy-dependent fits. More details about the SAID
program can be found in Refs.\cite{said} and \cite{said1}.

Following the formulation of Ref. \cite{SL96},
we write the resonant amplitude as
\begin{eqnarray}
\label{resampeq}
A^{N^*}_{m_f,m_i,\lambda_\gamma}({\bf q},{\bf k})
=\frac{1}{2\pi}\frac{M_N}{\sqrt{s}}\sqrt{\frac{q}{k}}
I^{N^*}_{m_f,m_\pi,\lambda_\gamma}({\bf q},{\bf k}) \nonumber
\end{eqnarray}
\begin{eqnarray}
I^{N^*}_{m_f,m_i,\lambda_\gamma}({\bf q},{\bf k})
= \sum_{J,M_J^{}}
\frac{{\cal M}_{N^*\to N'\pi}({\bf q};m_f^{};J,M_J^{})
{\cal M}_{\gamma N \to N^*}({\bf k};m_i^{},\lambda_\gamma^{};J,M_J^{})}
{\sqrt{s} - M_R^J + \frac{i}{2}\Gamma^J(s)},
\label{T:N*}
\end{eqnarray}
where $M^J_R$ is the mass of a $N^*$ with spin quantum numbers
$(J, M_J)$.
Here we neglect the effect due to the non-resonant mechanisms on the
$N^*$ decay amplitude and the shift of the resonance position.
Then the resonance mass $M_R^J$ and the $N^*$ decay amplitude
${\cal M}_{N^* \to \gamma N, \pi N}$ can be identified with the
predictions from a quark model,  as discussed 
in Refs.  \cite{SL96,YSAL00}.

We however do not have information about the total decay width
$\Gamma^J(s)$ for most of the $N^*$'s considered here.
For simplicity, we assume that its energy-dependence is similar to
the width of the $N^* \rightarrow \pi N$ decay within the oscillator
constituent quark model.
Following Ref. \cite{YSAL00} and neglecting the real part of the mass
shift, we then have
\begin{equation}
\Gamma^J (s) = \Gamma^J_0
\frac{\rho(q)}{\rho(q_0)}
\left( \frac{q}{q_{0}} \right)^{2L}
\exp\left[2 ({\bf q}_{0}^2 - {\bf q}^2)/\Lambda^2 \right],
\label{N*:decay}
\end{equation}
where $L$ is the orbital angular momentum of the
considered $\pi N$ state and
\begin{equation}
\rho(q) = \frac{qE_N(q) E_\pi(q)}{E_N(q) + E_\pi(q)}.
\end{equation}
In the above equations, $q(\equiv |{\bf q}|)$ is the pion momentum
at energy $\sqrt{s}$ while the on-shell momentum
$q_{0}$ is evaluated at $\sqrt{s} = M^J_R$. The energies are
defined by $E_N(q)=\sqrt{M_N^2+{\bf q}^2}$ and
$E_\pi(q)=\sqrt{M_\pi^2+{\bf q}^2}$.
Our choice of the total average width $\Gamma^J_0$ and cutoff parameter
$\Lambda$ for Eq. (\ref{N*:decay}) will be specified in Sec.~III.

By setting the photon momentum in the $z$-direction, the
$\gamma N \to N^*$ amplitude (${\cal M}_{\gamma N \to N^*}$) in Eq. (\ref{T:N*}) can be calculated
from the helicity amplitude $A_{\lambda}$ listed in 
Ref. \cite{Caps92} 
\begin{equation}
{\cal M}_{\gamma N \to N^*}({\bf k};m_i^{},\lambda_\gamma^{};J,M_J^{})
= \sqrt{2k} \, A_{M_J^{}}^{}
\delta_{M_J^{}, \lambda_\gamma^{} + m_i^{}}.
\label{photoN*}
\end{equation}
The $N^* \to \pi N$ amplitude (${\cal M}_{N^*\to N'\pi}$) takes the following form \cite{caps00}
\begin{eqnarray}
{\cal M}_{N^*\to N'\pi}({\bf q};m_f^{};J,M_J^{})
&=& 2\pi \sqrt{\frac{2M_R^J}{M_N^{}|{\bf q}_0^{}|}}
\sum_{ m_L^{} }
\langle L\, m_L^{}\, \frac{1}{2}, m_f | J\, m_J^{} \rangle
\nonumber \\ && \times
Y_{L m_L^{}}(\hat{q}) \, G(L)
\left( |{\bf q}|/|{\bf q}^{}_0| \right)^L f({\bf q},{\bf q}^{}_0),
\label{omegaN*}
\end{eqnarray}
where ${\bf q}_0$
is the $\pi$ meson momentum at $\sqrt{s} = M_R^J$.
Here we also include an extrapolation factor $f({\bf q},{\bf q}^{}_0) =
\exp[({\bf q}_0^2-{\bf q}^2)/\Lambda^2]$ similar to that for defining
the  width Eq.(5).

By using the information in Refs.\cite{Caps92,CR94,caps00}, we can get
helicity amplitude $A_\lambda$ of Eq.(7) and partial-wave transition
strength $G(L)$ of Eq.(8) predicted by Capstick and Roberts.
The resulting values for the considered 30 $N^*$ and
26 $\Delta^*$ are listed in Tables I-IV.

\section{RESULTS AND DISCUSSIONS}
\label{results}
 
Our first task is to calculate the resonant amplitude
defined by Eqs.~(\ref{resampeq})-(\ref{omegaN*}). 
With the helicity amplitude
$A_\lambda$ and partial-wave transition strength $G(L)$
given in Tables I-IV, this amplitude depends only on the choice of
the total width $\Gamma^J_0$ and the cutoff $\Lambda$ in Eqs.(5) and (8). 
For simplicity, we set $\Lambda$ = 650 MeV for all
resonances considered. This value is close to what was determined in
the investigation of $\Delta\rightarrow \pi N$\cite{SL96}. 
The widths of the known $N^*$ resonances listed in ~\cite{PDG98} 
range from 200-400 MeV.  We therefore take their
average value and set  $\Gamma^{J}_{0}= 300$ MeV
for all $N^*$ resonances listed in Tables I-II.
For the similar reason, we set $\Gamma^{J}_{0}=120$ MeV for
the considered $\Delta^*$ resonances listed in Tables III-IV. 
With these specifications, the
resonant amplitude Eq.(4) can be computed with no adjustable parameters.
Since the full amplitude $A({\bf q},{\bf k})$ is taken from SAID program, 
the non-resonant(background) amplitude $A^{(b.g.)}$ for
our investigation can then be fixed by  
\begin{eqnarray}
\label{nrbg_eq}
 A^{(b.g.)}_{mf,mi,\lambda_\gamma}({\bf q},{\bf k})
=A_{mf,mi,\lambda_\gamma}({\bf q},{\bf k}) 
- \sum_{N^*,\Delta^*} [A^{N^*}_{m_f,m_i,\lambda_\gamma}({\bf q},{\bf k})].
\end{eqnarray}

\begin{figure}
\centerline{\epsfig{file=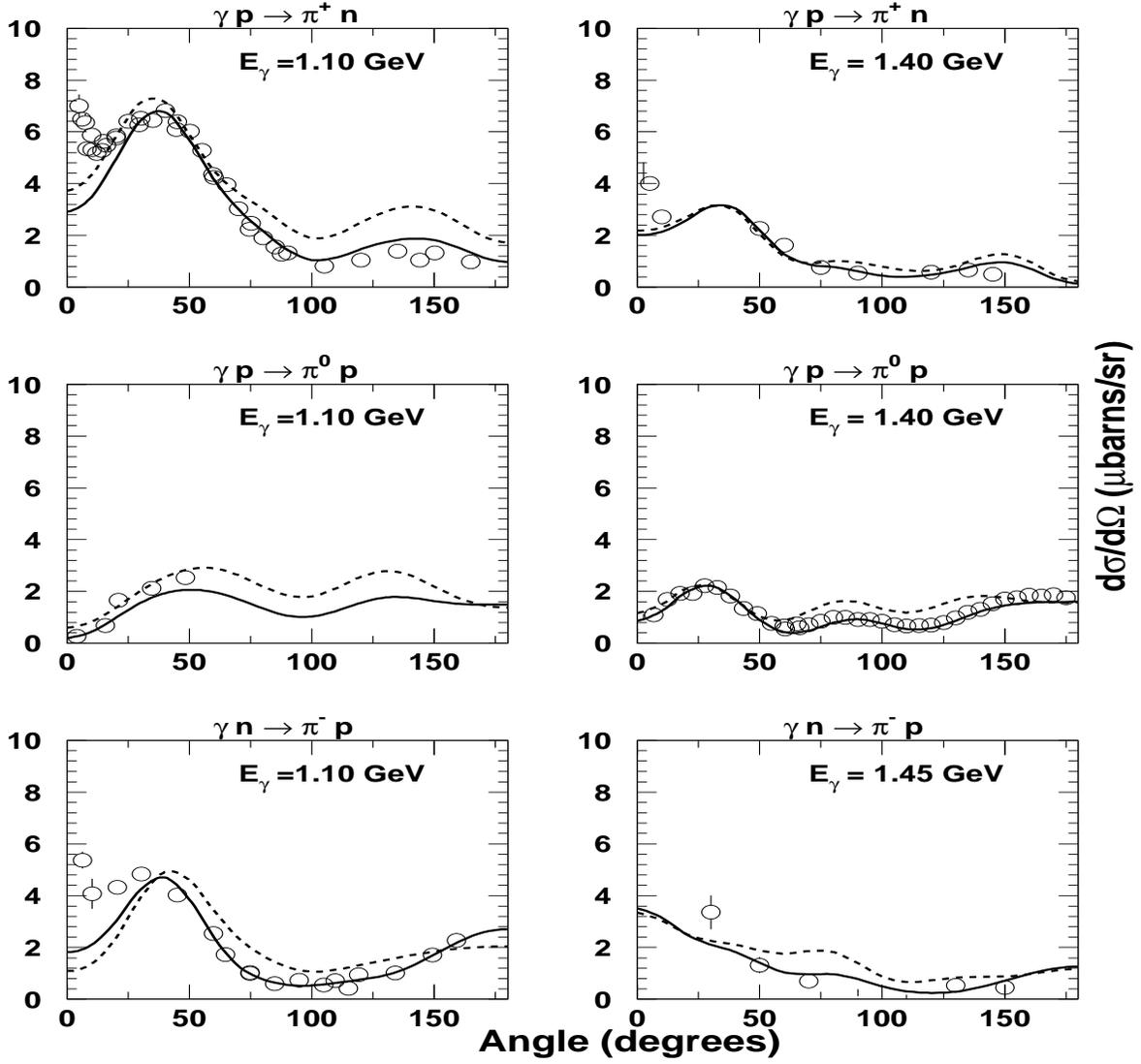,height=16.0cm,width=16.0cm}}
\caption[]{The differential cross sections 
of $\gamma N \rightarrow \pi N$ reactions calculated from
the background amplitude defined by Eq.(9)(dashed curves) and 
from the  SAID amplitude (solid curves) are compared. The data were obtained from Ref.~\cite{hepdata}}
\label{amp}
\end{figure}

With the background amplitude defined above, we then use Eq.(2) to
investigate the role of each resonance in determining the 
resonance amplitude.
In Fig.~\ref{amp}, we first observe that the differential cross
sections(solid curves) calculated from the employed SAID amplitude
 agree well with the available data. The dashed curves
are obtained when the  the effects due to all of the considered
56 resonance states are removed. We see that
the nucleon resonance effects on the unpolarized 
differential cross sections 
are not very pronounced.
Furthermore none of the considered resonances
listed in Tables I-IV dominate the difference between the 
full calculation(solid curves) and the background 
only calculation(dashed
curves). Obviously,
it is not easy to investigate the nucleon resonances by only
considering the unpolarized differential cross sections.
This is similar to the finding of Ref.\cite{oh2001} in
a investigation of $\omega$ photoproduction.

We now turn to reporting our investigations of the polarization
observables. Before we discussing the double polarization
observables, we show in Fig.~2 some typical results from our
calculations of single polarization observables.
As expected,
the effects due to nucleon resonances are much more visible here.
Hopefully, these predictions can be tested in the near future
when the data from the on-going experiments become available.

To be more specific in our investigation of the double polarization
observables, we focus on the most plausible
experiments on beam-target and beam-recoil polarizations defined
below
\begin{eqnarray}
\label{teq}
C^{BT}_{zy} = \frac{\sigma^{(r,y;U,U)} - \sigma^{(r,-y;U,U)}}{\sigma^{(r,y;U,U)} + \sigma^{(r,-y;U,U)}},
\end{eqnarray}

\begin{eqnarray}
\label{peq}
C^{BR}_{zy} = \frac{\sigma^{(r,U;y,U)} - 
\sigma^{(r,U;-y,U)}}{\sigma^{(r,U;y,U)} + \sigma^{(r,U;-y,U)}},
\end{eqnarray}
where $\sigma^{(B,T;R,M)}$ is the cross-section
 $\frac{d\sigma}{d\Omega}$ and the superscripts $(B,T;R,M)$ denote the
 polarizations of the beam, target; recoil and meson respectively. $r$ 
corresponds to a circularly polarized photon beam with 
helicity~+~1, $\pm~y$ represents the direction of the nucleon 
polarization, and U represents the 
unpolarized particle. The coordinate system is chosen such that
the incoming photon is in the z-direction and x-z plane is the
reaction plane. Thus y-direction is perpendicular to the reaction plane.

We have performed extensive calculations of the two double
 polarization observables, $C^{BT}_{zy}$ and $C^{BR}_{zy}$, in the 
energy region  1.6 GeV $<$ W $< $ 2.1 GeV.
As expected, the resonance effects are much more pronounced than what
can be observed in Fig.1 for the unpolarized differential cross sections or Fig.2. for the
single polarization observables. This is illustrated in Fig.3 for the beam-target
 polarization and Fig.4 for the beam-recoil polarization for two typical angles at 90 and 
120 degrees. 

Our next task is   
to explore the favorable kinematic regions where only one or just
a limited number
of nucleon resonances among the 56 considered resonance states will
dominant. We have found that at angles close to 120 degrees
the resonance effects on the  beam-recoil
polarization( Eq.(11)) are mainly due to the four-star
resonances $N^+_{9/2}(2345)$ of Table I and
$N^-_{7/2}(2090)$ and $N^-_{9/2}(2215)$ of table II,
and the two-star resonance $N^-_{3/2}(1960)$ of table II.
Furthermore the $N^-_{3/2}(1960)$ state has the largest effect
 among about 50 two-star and missing resonances in this higher
energy region.
 This is shown in Fig.5. We see that by adding only the two-star
$N^-_{3/2}(1960)$ to the calculation including
only background and three four-star resonances, the predicted 
beam-recoil polarizations are
shifted from the dashed curves to the dot-dashed curves which are
close to the results from full calculations including all
resonances listed in Tables I-IV. The investigation of the
two-star $N^-_{3/2}(1960)$ is clearly most favorable in the
$\gamma p \rightarrow \pi^+ n$ channel.

Our results suggest that data on the double polarization observables 
of pion photoproduction in 
the region 1.6~$\leq$~W~$\leq$2.1 GeV could be very useful in
investigating nucleon resonances.
The experiments on these observables would be complementary
to other similarly motivated ongoing experiments
on vector meson photoproduction and
kaon photoproduction.  Moreover, these double polarization
observables should be accessible at facilities such as Jefferson Lab 
and Spring-8.
To end, we like to mention that
some experimental data on double polarization 
observables have recently become available 
in the $\gamma~p~\rightarrow~\pi^0~p$ channel for W~$>$~1.6 GeV~\cite{krishni2001}, but
there is no data on the other two charged channels. 

\vspace{3cm}
This work was supported
by the U.S. Department of Energy under
contract number DE-FC02-94ER40818,  and also by
U.S. Department of
Energy, Nuclear Physics Division, contract No. W-31-109-ENG-38.

%%%%%%%%%%%%%%%%%  References  %%%%%%%%%%%%%%%%%%%%%%%

%%%%%%%%%%%%%%%%%  Tables  %%%%%%%%%%%%%%%%%%%%%%%

\begin{table}
\centering
\begin{tabular}{ccccccccc}
%%%%%%%%%%%%%%%%% spin 1/2
$N^*$& $M^J_R$ & $A_{1/2}$ & $A_{3/2}$ &
$G(1)$  & PDG \cite{PDG98} \\ \hline
$N\frac12^+$ & $1540$ & $4$ & --- & $20.3$ & 
$P_{11}(1440)^{\star\star\star\star}$ \\
$N\frac12^+$ & $1770$ & $13$ & --- & $4.2$ & 
$P_{11}(1770)^{\star\star\star}$ \\
$N\frac12^+$ & $1880$ & $0$   &  ---  & $2.7$ & \\
$N\frac12^+$ & $1975$ & $-12$ &  ---  & $2.0$ & \\
\hline
%%%%%%%%%%%%%%%%% spin 3/2
  & & & & $G(1)$  & \\ \hline
$N\frac32^+$ & $1795$ & $-11$ & $-31$ & $14.1$ & 
$P_{13}(1720)^{\star\star\star\star}$ \\
$N\frac32^+$ & $1870$ & $-2$  & $-15$ & $6.1$  &
 $P_{13} (1900)^{\star\star}$ \\
$N\frac32^+$ & $1910$ & $-21$ & $-27$ & $1.0$ & \\
$N\frac32^+$ & $1950$ & $-5$  & $2$   & $4.1$ & \\
$N\frac32^+$ & $2030$ & $-9$  & $15$  & $1.8$ &  \\
\hline
%%%%%%%%%%%%%%%%% spin 5/2
  & & & & $G(3)$  &  \\
\hline
$N\frac52^+$ & $1770$ & $-38$ & $56$  & $6.6$ & 
$F{15}(1680)^{\star\star\star\star}$ \\
$N\frac52^+$ & $1980$ & $-11$ & $-6$  & $1.3$ &  \\
$N\frac52^+$ & $1995$ & $-18$ & $1$   & $0.9$ &
 $F_{15} (2000)^{\star\star}$ \\
\hline
%%%%%%%%%%%%%%%%% spin 7/2
  & & & & $G(3)$  & \\ \hline
$N\frac72^+$ & $1980$ & $-1$  & $-2$  & $2.4$ &
 $F_{17} (1990)^{\star\star}$ \\
$N\frac72^+$ & $2390$ & $-14$ & $-11$& $4.9$ & \\
$N\frac72^+$ & $2410$ & $+1$ & $-1$& $0.4$ &  \\
\hline
%%%%%%%%%%%%%%%%% spin 9/2
  & & & & $G(5)$ & \\ \hline
$N\frac92^+$ & $2345$ & $-29$ & $+13$& $3.6$ &
 $H_{19} (2220)^{\star\star\star\star}$ \\
\end{tabular}
\caption{
Parameters for positive parity nucleon resonances from Refs.
\protect\cite{Caps92,CR94}.
The helicity amplitude $A_\lambda$ is given in unit of
$10^{-3}$ GeV$^{-1/2}$.
$G(L)$
is in unit of MeV$^{1/2}$.
The resonance mass $M^J_R$ is in unit of MeV.}
\label{Nstar1}
\end{table}
\begin{table}
\centering
\begin{tabular}{ccccccccc}
%%%%%%%%%%%%%%%%% spin 1/2
$N^*$& $M^J_R$ & $A_{1/2}$ & $A_{3/2}$ &
$G(0)$ & PDG \cite{PDG98} \\ \hline
$N\frac12^-$ & $1460$ & $+76$ & --- & $14.7$ & 
$S_{11}(1535)^{\star\star\star\star}$ \\
$N\frac12^-$ & $1535$ & $+54$ & --- & $12.2$ &
$S_{11}(1650)^{\star\star\star\star}$  \\
$N\frac12^-$ & $1945$ & $+12$ & --- & $5.7$ &  \\
$N\frac12^-$ & $2030$ & $+20$ & --- & $3.7$ &  \\
\hline
%%%%%%%%%%%%%%%%% spin 3/2
 & & & & $G(2)$ & \\ \hline
$N\frac32^-$ & $1495$ & $-15$ & $134$ & $8.6$ &
$D_{13}(1520)^{\star\star\star\star}$  \\
$N\frac32^-$ & $1625$ & $-33$ & $-3$ & $5.8$ &
$D_{13}(1700)^{\star\star\star}$  \\
$N\frac32^-$ & $1960$ & $+36$ & $-43$ & $8.2$ &
$D_{13} (2080)^{\star\star}$ \\
$N\frac32^-$ & $2055$ & $+16$ & $0$ & $6.2$ &  \\
$N\frac32^-$ & $2095$ & $-9$ & $-14$ & $0.2$ &  \\
\hline
%%%%%%%%%%%%%%%%% spin 5/2
 & & & & $G(2)$ & \\ \hline
$N\frac52^-$ & $1630$ & $2$ & $3$ & $5.3$ &
$D_{15}(1675)^{\star\star\star}$ \\
$N\frac52^-$ & $2080$ & $-3$ & $-14$ & $5.1$ & \\
$N\frac52^-$ & $2095$ & $-2$ & $-6$ & $5.2$ &
 $D_{15} (2200)^{\star\star}$ \\
\hline
%%%%%%%%%%%%%%%%% spin 7/2
 & & & & $G(4)$ & \\ \hline
$N\frac72^-$ & $2090$ & $-34$ & $+28$ & $6.9$ &
 $G_{17} (2190)^{\star\star\star\star}$ \\
$N\frac72^-$ & $2205$ & $-16$ & $+4$ & $4.0$ &  \\
\hline
%%%%%%%%%%%%%%%%% spin 9/2
 & & & & $G(4)$  & \\ \hline
$N\frac92^-$ & $2215$ & $0$ & $+1$ & $2.5$ &
 $G_{19} (2250)^{\star\star\star\star}$ \\
\end{tabular}
\caption{
Parameters for negative parity nucleon resonances from Refs.
\protect\cite{Caps92,CR94}.
The units are the same as in Table \ref{Nstar1}.}
\label{Nstar2}
\end{table}

\begin{table}
\centering
\begin{tabular}{ccccccccc}
%%%%%%%%%%%%%%%%% spin 1/2
$\Delta^*$& $M^J_R$ & $A_{1/2}$ & $A_{3/2}$ &
$G(1)$  & PDG \cite{PDG98} \\ \hline
$\Delta\frac12^+$ & $1835$ & $-31$ & --- & $3.9$ & \\
$\Delta\frac12^+$ & $1875$ & $-8$ & --- & $9.4$ & 
$P_{31}(1910)^{\star\star\star\star}$\\
\hline
%%%%%%%%%%%%%%%%% spin 3/2
  & & & & $G(1)$  & \\ \hline
$\Delta\frac32^+$ & $1232$ & $108$ & $186$ & $10.4$ &
$P_{33}(1232)^{\star\star\star\star}$ \\
$\Delta\frac32^+$ & $1795$ & $30$  & $51$ & $8.7$  &
 $P_{33} (1920)^{\star\star}$ \\
$\Delta\frac32^+$ & $1915$ & $13$ & $14$ & $4.2$ & \\
$\Delta\frac32^+$ & $1985$ & $6$  & $3$   & $3.3$ & \\
\hline
%%%%%%%%%%%%%%%%% spin 5/2
  & & & & $G(3)$  &  \\
\hline
$\Delta\frac52^+$ & $1910$ & $26$ & $-1$  & $3.4$ &
$F_{15}(1905)^{\star\star\star\star}$  \\
$\Delta\frac52^+$ & $1990$ & $-10$ & $-28$  & $1.2$ & 
 $F_{35}(2000)^{\star\star}$ \\
\hline
%%%%%%%%%%%%%%%%% spin 7/2
  & & & & $G(3)$  & \\ \hline
$\Delta\frac72^+$ & $1940$ & $-33$  & $-42$  & $7.1$ &
 $F_{37}(1950)^{\star\star\star\star}$ \\
$\Delta\frac72^+$ & $2370$ & $-33$ & $-42$& $1.5$ & 
$F_{37}(1950)^{\star\star\star\star}$ \\
$\Delta\frac72^+$ & $2460$ & $24$ & $30$& $1.1$ & 
 \\
\hline
%%%%%%%%%%%%%%%%% spin 9/2
  & & & & $G(5)$ & \\ \hline
$\Delta\frac92^+$ & $2420$ & $0$ & $0$& $1.2$ &
 $H_{39}$\\
$\Delta\frac92^+$ & $2505$ & $0$ & $0$& $0.4$ & \\
\end{tabular}
\caption{
Parameters for positive parity Delta resonances from Refs.
\protect\cite{Caps92,CR94}.
The helicity amplitude $A_\lambda$ is given in unit of
$10^{-3}$ GeV$^{-1/2}$.
$G(L)$
is in unit of MeV$^{1/2}$.
The resonance mass $M^J_R$ is in unit of MeV.}
\label{Deltastar1}
\end{table}
\begin{table}
\centering
\begin{tabular}{ccccccccc}
%%%%%%%%%%%%%%%%% spin 1/2
$\Delta^*$& $M^J_R$ & $A_{1/2}$ & $A_{3/2}$ &
$G(0)$ & PDG \cite{PDG98} \\ \hline
$\Delta\frac12^-$ & $1555$ & $81$ & --- & $5.1$ &  
$S_{31}(1620)^{\star\star\star\star}$\\
$\Delta\frac12^-$ & $2035$ & $20$ & --- & $1.2$ &
$S_{31}(1900)^{\star\star}$  \\
$\Delta\frac12^-$ & $2140$ & $4$ & --- & $3.1$ &  \\
\hline
%%%%%%%%%%%%%%%%% spin 3/2
 & & & & $G(2)$ & \\ \hline
$\Delta\frac32^-$ & $1620$ & $82$ & $68$ & $4.9$ & 
$D_{33}(1700)^{\star\star\star\star}$ \\
$\Delta\frac32^-$ & $2080$ & $-20$ & $-6$ & $2.2$   \\
$\Delta\frac32^-$ & $2145$ & $0$ & $10$ & $2.2$ & \\
\hline
%%%%%%%%%%%%%%%%% spin 5/2
 & & & & $G(2)$ & \\ \hline
$\Delta\frac52^-$ & $2155$ & $11$ & $19$ & $5.2$ & 
$D_{35}(1930)^{\star\star\star}$ \\
$\Delta\frac52^-$ & $2165$ & $0$ & $0$ & $0.6$ & \\
$\Delta\frac52^-$ & $2265$ & $0$ & $0$ & $2.4$ & \\
$\Delta\frac52^-$ & $2325$ & $0$ & $0$ & $0.1$ & \\
\hline
%%%%%%%%%%%%%%%%% spin 7/2
 & & & & $G(4)$ & \\ \hline
$\Delta\frac72^-$ & $2230$ & $14$ & $-4$ & $2.1$ &
 $G_{37}$ \\
$\Delta\frac72^-$ & $2295$ & $0$ & $0$ & $1.8$ &  \\
\hline
%%%%%%%%%%%%%%%%% spin 9/2
 & & & & $G(4)$  & \\ \hline
$\Delta\frac92^-$ & $2295$ & $-14$ & $-17$ & $4.8$ &
 $G_{39}$ \\
\end{tabular}
\caption{
Parameters for negative parity Delta resonances from Refs.
\protect\cite{Caps92,CR94}.
The units are the same as in Table \ref{Deltastar1}.}
\label{Deltastar2}
\end{table}

\newpage

\begin{figure}
\centerline{\epsfig{file=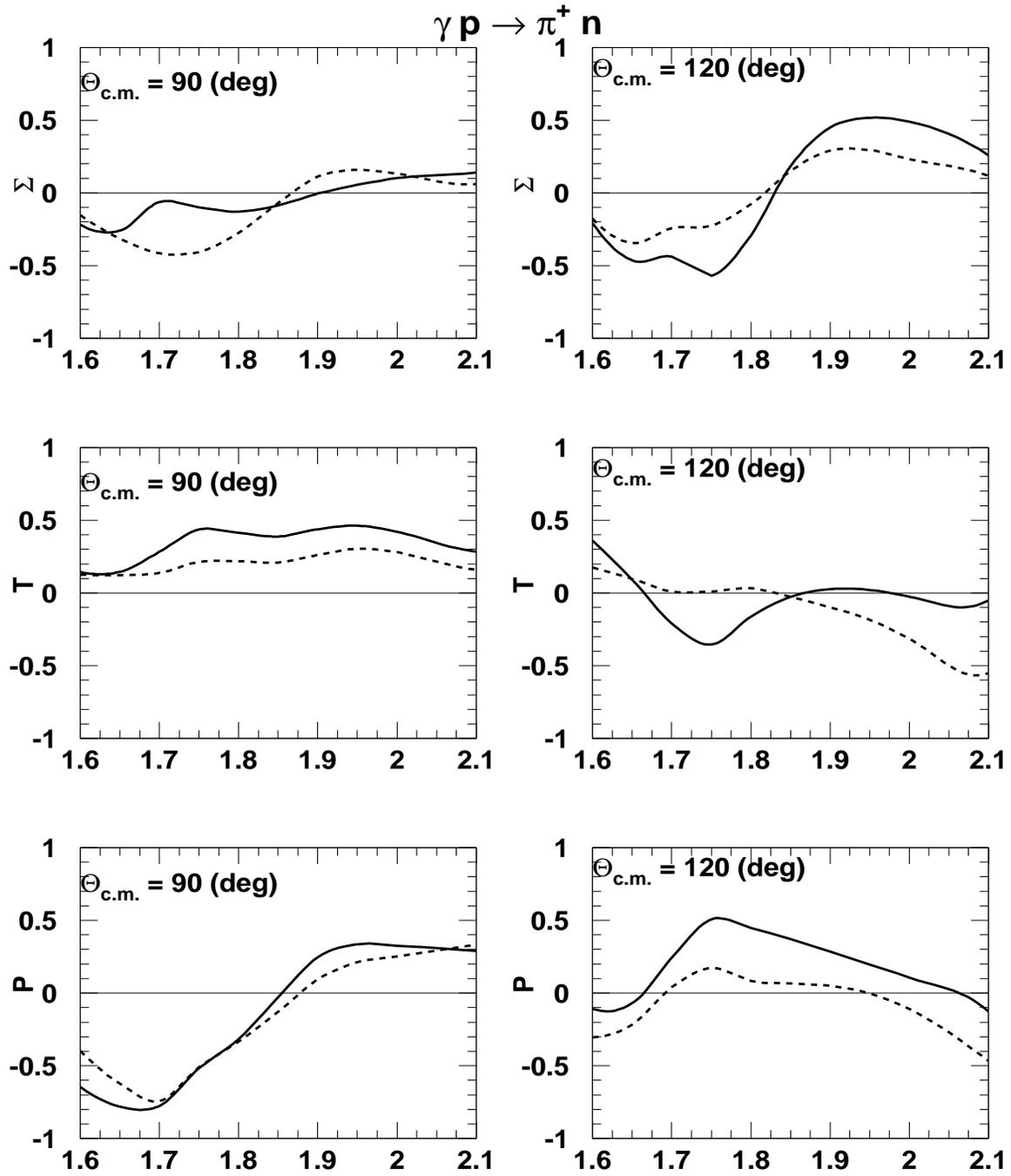,height=20.0cm,width=16.0cm}}
\caption{The single polarization observables, $\gamma$ asymmetry ($\Sigma$), the target asymmetry (T) and the recoil asymmetry (P) at c.m. angle 90 and 
120 degrees. The solid(dashed) curves are obtained from using the SAID amplitude (the background amplitude defined by Eq.(9)).}
\label{singpol}
\end{figure}

\begin{figure}
\centerline{\epsfig{file=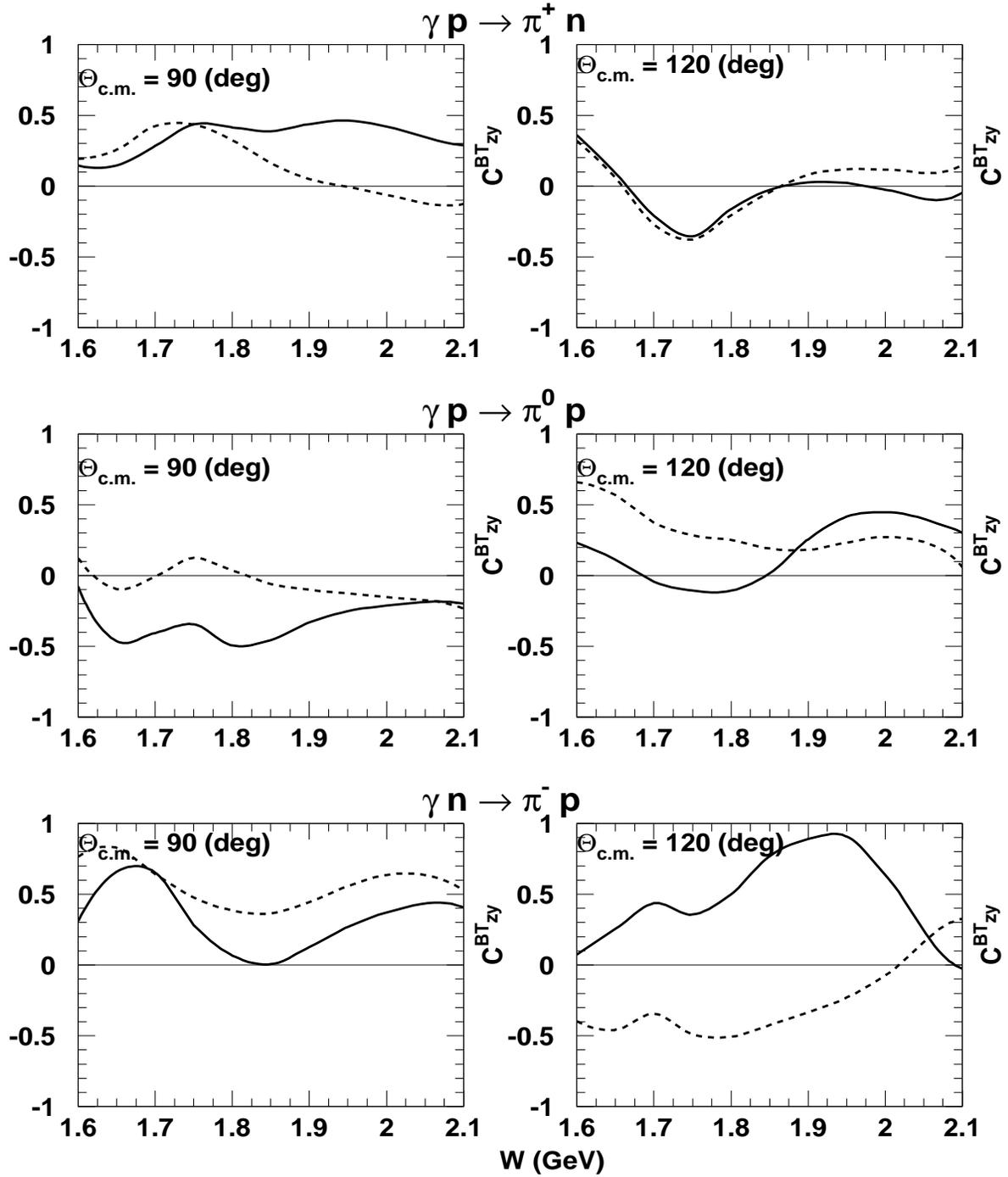,height=20.0cm,width=16.0cm}}
\caption{The beam-target polarization C$^{BT}_{zy}$ with 
photon helicity = +1 at c.m. angle 90 and 120 degrees.
The solid(dashed) curves are obtained from using the
 SAID amplitude (the background amplitude
defined by Eq.(9)).}
\label{unknt}
\end{figure}

\begin{figure}
\centerline{\epsfig{file=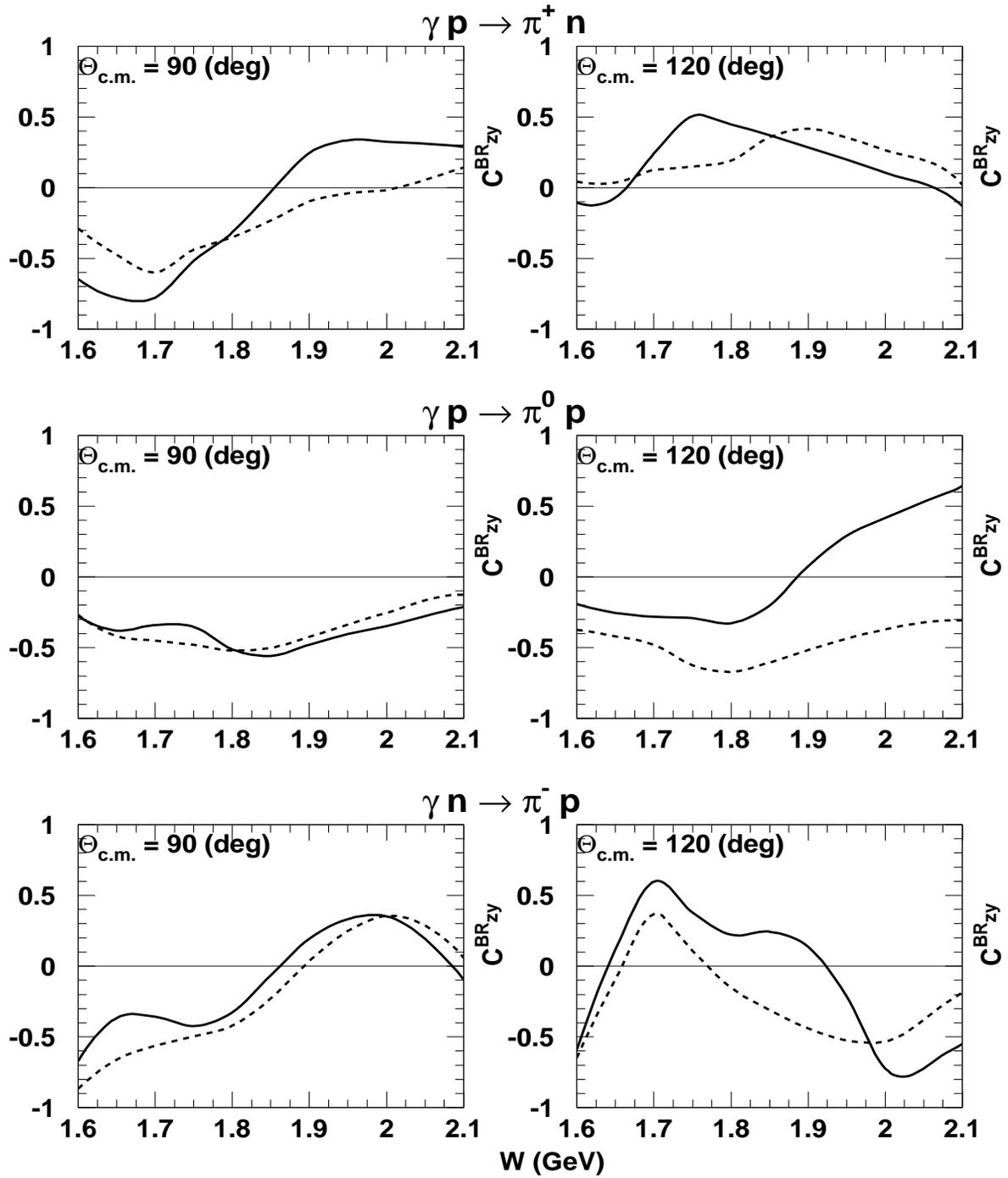,height=20.0cm,width=16.0cm}}
\caption{Same as Fig.3 except for the beam-recoil polarization 
C$^{BR}_{zy}$.}
\label{unknp}
\end{figure}
\begin{figure}
\centerline{\epsfig{file=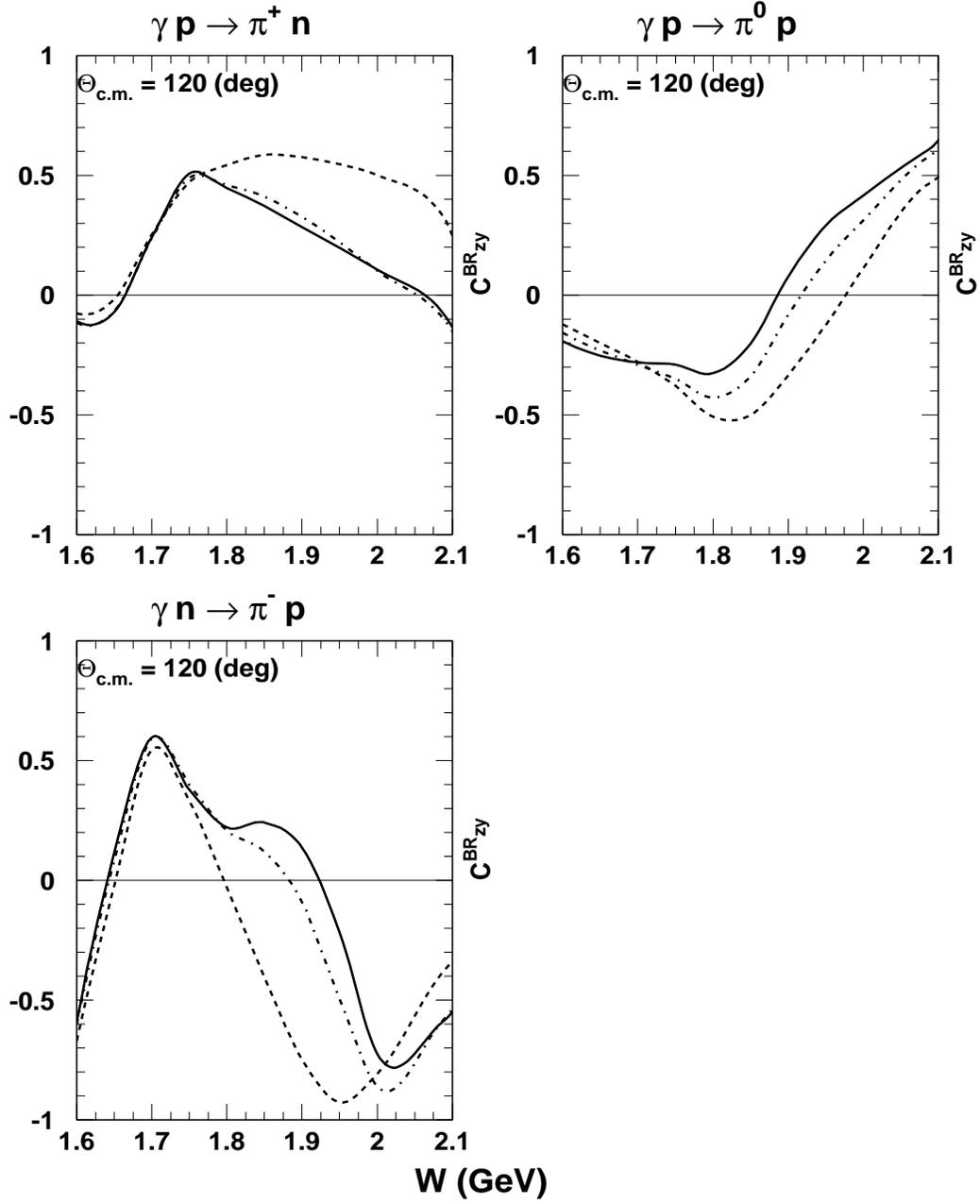,height=18.0cm,width=14.0cm}}
\caption{The beam-recoil polarization C$^{BR}_{zy}$ with
photon helicity = +1 at c.m. angle 120 degree. 
The solid curves are from the  SAID amplitude.
The dashed curves are from the 
background amplitude + four-star resonances.
The dot-dashed curves are from the background amplitude 
+ four-star resonances + N$^{-}_{3/2}$(1960).}
\label{n1960p}
\end{figure}

\end{document}